\documentclass[nofootinbib,prd,aps,onecolumn,preprintnumbers,amsmath,amssymb,superscriptaddress]{revtex4}
\usepackage{amsmath}
\usepackage{amssymb}
\usepackage{graphicx}
\usepackage{subfigure}
\usepackage{color}
\usepackage[colorlinks,linkcolor=magenta,anchorcolor=blue,citecolor=green]{hyperref}
\usepackage{ulem}
\usepackage{pifont}
\usepackage{makecell}
\usepackage{url}

\begin{document}
	
\title{Generation of primordial black holes from an inflation model with modified dispersion relation}

\author{Taotao Qiu}
\email{qiutt@hust.edu.cn}
\affiliation{School of Physics, Huazhong University of Science and Technology\\
Wuhan, 430074, China}
\author{Wenyi Wang}
\email{wangwy@mails.ccnu.edu.cn(corresponding author)}
\affiliation{Institute of Astrophysics, Central China Normal University, Wuhan 430079 ,China}
\author{Ruifeng Zheng}
\email{zrf2022@stu2022.jnu.edu.cn}
\affiliation{Department of Physics and Siyuan Laboratory, Jinan University\\ 
Guangzhou 510632, China}
	
\begin{abstract}
A primordial black hole (PBH) is interesting to people for its ability of explaining dark matter as well as supermassive astrophysical objects. In the normal inflation scenario, the generation of PBHs usually requires an enhanced power spectrum of scalar perturbation at the end of inflation era, which is expected when the dispersion relation of the inflaton field gets modified. In this work, we study a kind of inflation model called {the \it Dirac-Born-Infeld-inspired nonminimal kinetic coupling (DINKIC)} model, where the dispersion relation is modified by a square root existing in the field Lagrangian. We discuss the enhancement of scalar power spectrum due to the modified dispersion relation, as well as the abundance of PBHs produced by the Press-Schechter collapse mechanism. We also discuss the formation of scalar-induced gravitational waves by linear scalar perturbations.

\end{abstract}
	
\maketitle
	
\section{introduction}\label{Sec: introduction}
	
Primordial black holes (PBHs) have been drawing attentions of more and more astrophysicists and cosmologists. Unlike the formation of astrophysical black holes, PBHs are not formed by the collapse of stars, but by the gravitational collapse of local high-density regions in the early Universe, thus PBHs have much broader mass range than astrophysical ones. Therefore, it cannot only act as dark matter whose identity has not been confirmed yet, but also an interesting candidate of the supermassive black hole, which seems impossible to be astrophysical because of the lack of formation time. First initiated by Zeldovich and Novikov \cite{Zeldovich:1967lct} in the 1960s, and put forward by Hawking and Carr in the 1970s~\cite{Hawking:1971ei,Carr:1974nx}, PBHs have been widely studied, see e.g. ~\cite{Khlopov:2008qy,Belotsky:2014kca,Sasaki:2018dmp,Belotsky:2018wph,Yuan:2021qgz,Villanueva-Domingo:2021spv,Escriva:2022duf} and the references therein. Moreover, there are also a lot of efforts putting various constraints on PBHs, such as from gravitational lensing~\cite{EROS-2:2006ryy,Niikura:2017zjd,Jung:2017flg, Niikura:2019kqi}, cosmic microwave background (CMB) and big bang nucleosynthesis (BBN)~\cite{Carr:2009jm, Serpico:2020ehh, Acharya:2020jbv}, gamma-ray emission~\cite{Carr:2009jm,Barnacka:2012bm, Laha:2019ssq, Dasgupta:2019cae, Laha:2020ivk, Cai:2020fnq, Tan:2022lbm}, compact objects~\cite{Graham:2015apa, Capela:2013yf, Lu:2019ktw}, gravitational waves~\cite{Chen:2019xse, Wong:2020yig,Kimura:2021sqz, Kavanagh:2018ggo,Wang:2022nml}, large-scale structure (LSSs)~\cite{Carr:2018rid} and so on; see \cite{Carr:2020gox} for a review.

As has been demonstrated in the literature, PBHs can be generated in inflation scenario. 
During the inflation era, the Universe expands dramatically over a short period of time, while the quantum fluctuations in the vacuum of the inflation field will be stretched out of the horizon and become classical perturbations. In small scales, if the power spectrum of the cosmological perturbations has large peaks, it will lead to large inhomogeneities in the energy distribution of the Universe. After the perturbation reenters the horizon, PBHs will form in regions of high energy density due to gravitational collapse~\cite{Riotto:2002yw}. 
To be precise, in order to effectively form the PBHs, it is necessary to enhance the amplitude of the power spectrum on small scales to the order of $10^{-2}$~\cite{Motohashi:2017kbs}, while on the CMB scale, it is constrained to $10^{-9}$ by the observations~\cite{Planck:2018jri}. 
There are many ways to enhance the power spectrum, such as selection of scalar potentials with special features~\cite{Cai:2019bmk,Ketov:2019mfc,
Drees:2011yz,Garcia-Bellido:2017mdw,Di:2017ndc,Gao:2018pvq,Cheng:2018yyr,Xu:2019bdp,Lin:2020goi,Ozsoy:2020kat,Kawai:2021edk,Kawai:2021bye,Solbi:2021wbo,Zheng:2021vda,Gangopadhyay:2021kmf,Ashoorioon:2020hln,Ashoorioon:2022raz,Karam:2022nym}, multifield inflation models~\cite{Garcia-Bellido:1996mdl,Bugaev:2011wy,Clesse:2015wea,Kawasaki:2012wr,Ahmed:2021ucx,Kawai:2022emp}, sound speed resonance~\cite{Cai:2018tuh,Chen:2019zza} and so on~\cite{Lin:2012gs,Pi:2017gih,Choudhury:2013woa,Fu:2019ttf,Arya:2019wck,Martin:2019nuw,Ashoorioon:2019xqc,Martin:2020fgl,Choudhury:2023vuj}.

Recently, there are works discussing about generating large scalar power spectrum by suppressing the sound speed of the inflaton field to a very tiny value, namely $c_s\ll 1$~\cite{Ballesteros:2021fsp,Gorji:2021isn,Zhai:2022mpi}. One can see from the expression of the scalar power spectrum, $P_\zeta\sim H^2/(\epsilon c_s)$ that, such an approach is parallel to that of suppressing the slow-roll parameter, as what has been done in the ultra-slow-roll inflation models \cite{Di:2017ndc,Motohashi:2017kbs,Ballesteros:2017fsr,Zheng:2021vda}, albeit the latter violates the slow-roll condition. Moreover, in these works, the suppression of $c_s$ was realized accompanied by a higher order term, in order not to violate the consistency requirement and lead to strong coupling \cite{Ballesteros:2021fsp,Gorji:2021isn,Ballesteros:2018wlw}. In \cite{Ballesteros:2021fsp}, such a realization was discussed in the general effective field theory language, while in \cite{Gorji:2021isn}, people used the ghost inflation with a higher-order corrected dispersion relation as a specific example.
	
On the other hand, the modified dispersion relation (with higher-order term) can be naturally generated in inflation models which have nonlinear kinetic terms, such as the "{\it DBI-inspired non-minimal kinetic coupling"
(DINKIC)} model proposed by one of the authors in 2015 \cite{Qiu:2015aha}. The nonlinearity in this model is due to the fact that the nonminimal kinetic coupling term resides inside the square root in the field Lagrangian. As a result, there is an additional term proportional to $k^4$ besides the normal dispersion relation: $\omega^2=c_s^2k^2$.  In this work, we discuss the generation of PBHs in the framework of this model. While in the original paper \cite{Qiu:2015aha}, we set the $c_s^2$ to be constant, in this work we make it vary, which gives rise to a modified dispersion relation: as the evolution goes, the $k^2$ term dominates first, and the $k^4$ dominates later. In such a case, we calculate the scalar perturbations in order to obtain a large power spectrum on small scales. We analyze the possibility of the formation of PBHs which can act as a large amount of dark matter, and confront our results to the constraints of current observations.
We also discuss the scalar-induced gravitational waves (SIGWs) generated in this model~\cite{Ananda:2006af,Baumann:2007zm,Saito:2008jc,Saito:2009jt,Domenech:2021ztg}.

This paper is organized as follows. In Sec.~\ref{DBI}, we briefly review the DINKIC inflation model, which contains a correction term of $k^{4}$ in the dispersion relation. In Sec.~\ref{model building}, we set up our model with varying sound speed, and present the information for background quantities. In Sec.~\ref{analytical solution}, we calculate the evolution of the perturbation so as to obtain the scalar power spectrum, and analyze the conditions for enhancing the power spectrum to $10^{-2}$. In Sec.~\ref{PBHs}, we calculate the PBH abundance produced in the model, and constrain our results with the current observations.  In Sec.~\ref{SIGWs}, we discuss the production of SIGWs in our model. Section~\ref{Conclusion} is devoted to the conclusions and discussions.
	
\section{DINKIC inflation model}\label{DBI} 
	
The original DINKIC inflation model proposed in Ref.~\cite{Qiu:2015aha} has the following action:
\begin{align} 
\label{Eq: action}
S=\int d^{4}x\sqrt{-g}\left[\frac{R}{2\kappa^{2}}-\frac{1}{f(\phi)}(\sqrt{\mathcal{D}}-1)-V(\phi)\right]
,
\end{align}
where we have defined
\begin{align}
\mathcal{D}&\equiv1-2\alpha f(\phi)X+2\beta f(\phi)\widetilde{X}
,\\
X&\equiv-\frac{1}{2}g^{\mu\nu}\nabla_{\mu}\phi\nabla_{\nu}\phi
,\\
\widetilde{X}&\equiv-\frac{1}{2M^{2}}G^{\mu\nu}\nabla_{\mu}\phi\nabla_{\nu}\phi.
\end{align}
Here $\alpha$ and $\beta$ are constants, $M$ is the scale of nonminimal kinetic coupling, while $M_{pl}=\kappa^{-1}$ is the Planck scale, and the function
$f(\phi)$ is the (squared) warp factor of the AdS-like throat. Note that the second kinetic term, $\widetilde{X}$, belongs to the generalized scalar-tensor theory action, which possesses nice properties such as violating the null energy condition without having ghosts \cite{Deffayet:2011gz, Kobayashi:2019hrl}. Therefore, it is interesting to extend it to the nonlinear action as well. One extension is to have the field's  action like that of the DBI field, which has strong motivations from string theory \cite{Polchinski:1998rq, Gerasimov:2000zp, Kutasov:2000qp, Kutasov:2000aq}. Note that the above action is related to Fab 5 theory proposed years ago \cite{Appleby:2012rx,Linder:2013zoa}.

Under the flat FLRW metric ($g_{\mu\nu}=diag\{-1,a^2(t),a^2(t),a^2(t)\}$), 
one can vary the action with respect to the field $\phi$ to get the equation of motion for $\phi$:
	\begin{align}\label{eom}
		\frac{f_{\phi}(\sqrt{\mathcal{D}}-1)^{2}}{2f^{2}\sqrt{\mathcal{D}}}+\frac{3\beta H^{2}-\alpha}{\sqrt{\mathcal{D}}}\ddot{\phi}+\frac{2\beta\dot{H}+3\beta H^{2}-\alpha}{\sqrt{\mathcal{D}}}3H\dot{\phi}-\frac{3\beta H^{2}-\alpha}{2\mathcal{D}^{3/2}}\dot{\mathcal{D}}\dot{\phi}-V_{\phi}=0
		~, 
	\end{align}
while the energy density $\rho$ and pressure $p$ are given by
	\begin{align}
		\rho &= \frac{(\sqrt{\cal D}-1)}{f(\phi)}+V(\phi)+\frac{\alpha\dot{\phi}^{2}}{\sqrt{\mathcal{D}}}+\frac{6\beta H^{2}\dot{\phi}^{2}}{M^2\sqrt{\mathcal{D}}},\\
		\label{p}
		p &= -\frac{(\sqrt{\cal D}-1)}{f(\phi)}-V(\phi)-\frac{3\beta H^{2}\dot{\phi}^{2}}{M^2\sqrt{\mathcal{D}}}-\left(\frac{\beta H\dot{\phi}^{2}}{M^2\sqrt{\mathcal{D}}}\right)^{.}.
	\end{align}
	
In order to analyze the perturbations generated by the model, we use the ADM formalism, in which the perturbed action up to the second order becomes
	\begin{align}\label{perturbations}
		S^{c}_{2}\approx\frac{1}{2\kappa^{2}}\int d^{4}xa^{3}\Big[6\frac{x_\beta}{\mathcal{D}}\dot{\zeta}^{2}-\frac{2\epsilon}{a^{2}}(\partial\zeta)^{2}+\frac{16x_\beta^{4}y}{a^{4}H^{2}}(\partial^{2}\zeta)^{2}\Big]~,
	\end{align}
where we define several dimensionless variables,
	\begin{align}\label{xandy}
		x_{\beta} = \frac{\kappa^{2}\beta\dot{\phi}^{2}}{2M^2\sqrt{\mathcal{D}}},	\quad y = \frac{f(\phi)M_{p}^2H^{2}}{\sqrt{\mathcal{D}}},
	\end{align}
and the $\epsilon\equiv-\dot H/H^2$ is the slow-roll parameter. One can see from Eq.~\eqref{perturbations} that, different from usual generalized scalar-tensor theory, an additional higher-order spatial derivative term appears, which is due to the nonlinearity of the action. 
From Eq.~\eqref{perturbations}, we can easily get the perturbation equation:
	\begin{align}\label{perturbeomscalar2}
		u^{\prime\prime}+c_s^2k^2\left[1+\left(\frac{k}{k_{c}}\right)^2\right]u-\frac{z^{\prime\prime}}{z}u=0,
	\end{align}
where the prime denotes derivative with respect to conformal time $d\tau\equiv a^{-1}(t) dt$; see Ref.~\cite{Qiu:2015aha} for more details. In the above equation, we define $u\equiv z\zeta$, $z\equiv a\sqrt{3x_\beta/{\cal D}}$, $c_{s}^{2}=\epsilon{\cal D}/3x_\beta$, and the critical scale:
\begin{align}
\label{kc}
    k_{c}\equiv aH\sqrt{\frac{\epsilon{\cal D}}{8x^{4}_{\beta}y}}=\frac{aHc_s}{\sqrt{\gamma}},
\end{align}
where $\gamma\equiv 8 x^3_\beta y/3$. Therefore we have the dispersion relation as
\begin{align}\label{dispersion relation}
    \omega^2=c_s^2k^2+c_s^2k_c^{-2}k^4.
\end{align}
The dispersion relation above indicates that the fluctuation modes of the inflation field can be divided into two cases, namely $k<k_c$ and $k>k_c$. In the first case, the first term in Eq. \eqref{dispersion relation} dominates over the second term, and the approximate dispersion relation approaches $\omega\sim k$. In the second case, the second term dominates over the first one, and it becomes $\omega\sim k^2$. 
The similar dispersion relation also appears in ghost inflation~\cite{Arkani-Hamed:2003juy} whose background has a timelike scalar field $\phi = M^{2} t$, and has been recently studied in EFT inflation~\cite{Ballesteros:2021fsp,Gorji:2021isn}.

\begin{figure*}
		\centering
		\includegraphics[height=6cm,width=8cm]{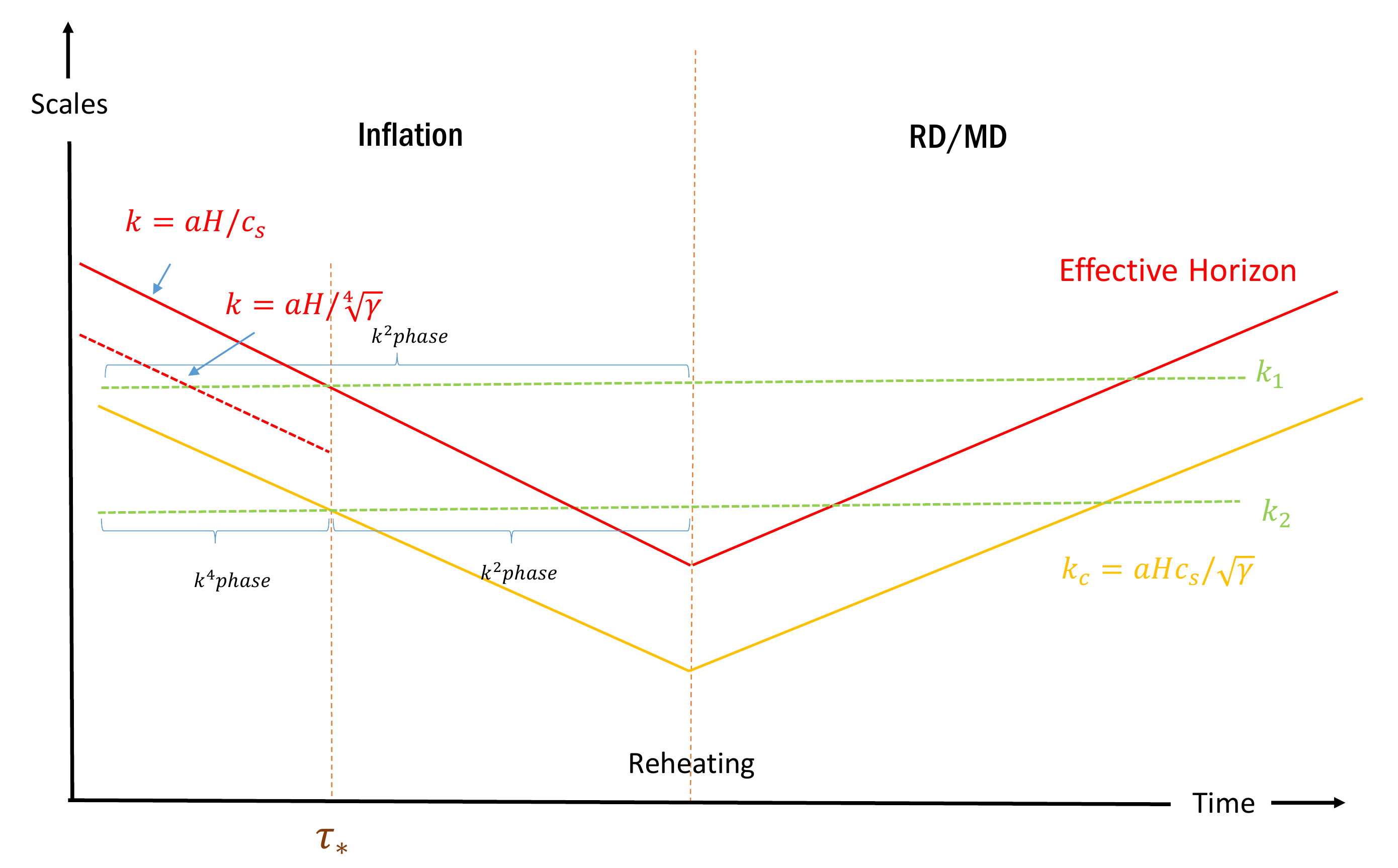}
        \includegraphics[height=6cm,width=8cm]{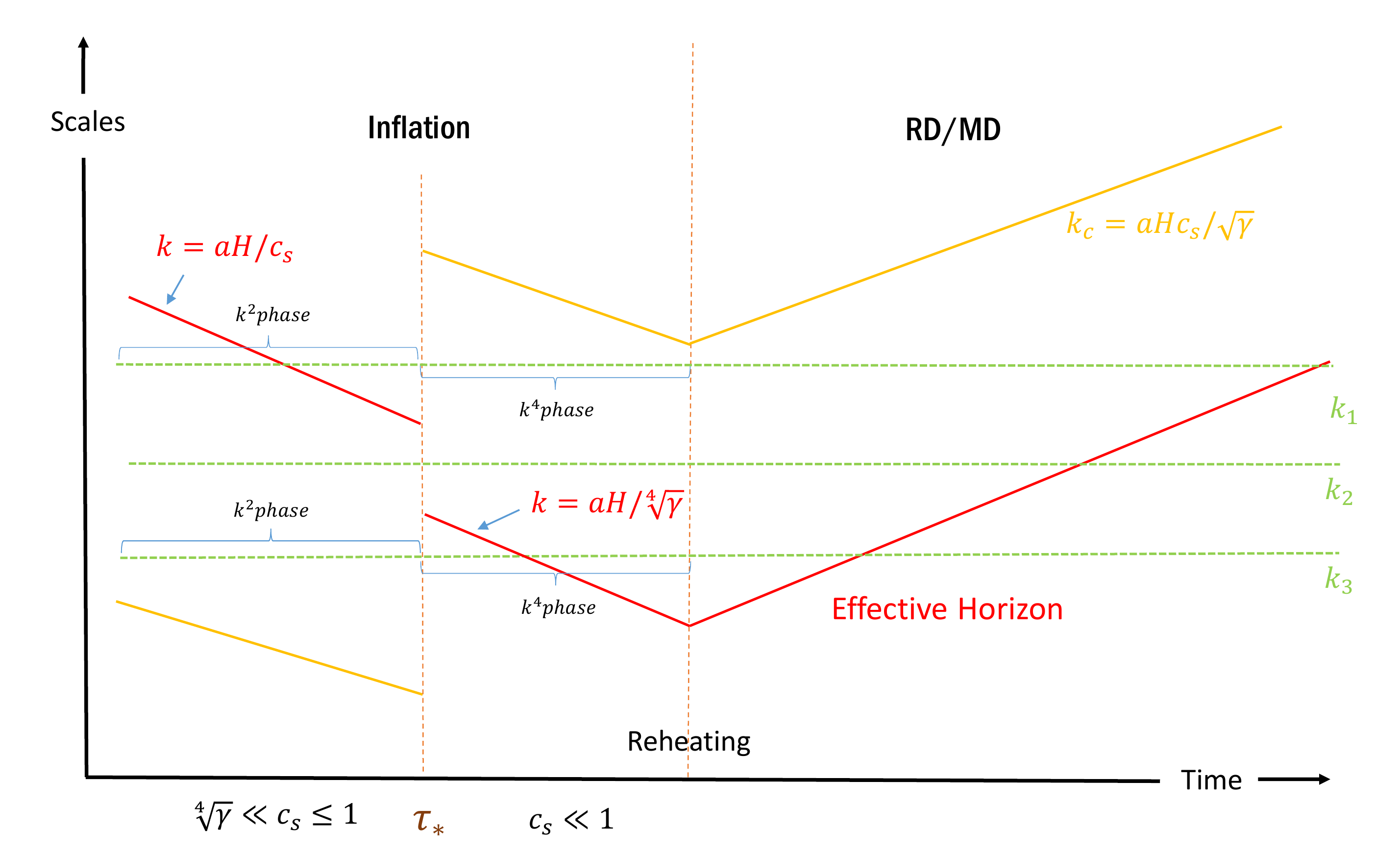}
		\caption{The sketch plots of the evolution of fluctuation modes for the DINKIC inflation model in \cite{Qiu:2015aha} (left panel) and this paper (right panel). The green lines: the conformal  wavelengths $1/k$ for different fluctuation modes; the yellow lines: the conformal “critical wavelength” $1/k_c$; the red lines: the “effective horizo” which presents the effects of the "effective potential" $z^{\prime\prime}/z$ in Eq. \eqref{perturbeomscalar2}, and it depends on which term in Eq. \eqref{dispersion relation} will dominate over the other. One can see that, while in the original model the effective horizon and the critical wavelengths are continuous, in this work both are steplike, which is useful to create the peak in the scalar power spectrum in small scales.} 
		\label{shiyitu}
\end{figure*}

In Ref. \cite{Qiu:2015aha}, we assume the parameters such as $x_\beta$, $y$, $\epsilon$ and $c_s^2$ are slow varying, so that $k_c\sim aH$. We draw the evolution of the “critical wavelength” $1/k_c$ as well as the wavelength of fluctuation mode with arbitrary wave number $k$ for this case in the left panel of Fig.~\ref{shiyitu}.
One can see that, for the modes with $k<k_c$ initially, it will keep so untill the end of inflation, so the $k^4$ term will be subdominant all the time. However, for the modes with $k>k_c$
initially, it will evolve untill $k$ becomes smaller than $k_c$ at a later time. According to the analysis in the previous section, the dispersion relation is dominated first by the $k^4$ term, then the $k^2$ term. 

We also draw the effective horizon $l_{eff}$ such that, when $1/k<l_{eff}$, the term with $k$ dominates over the “effective potential” term containing $z^{\prime\prime}/z$ (subhorizon), and vice versa (superhorizon). This determines the functional form of the solution $u$, as will be seen later. Thus, the $l_{eff}$ also depends on which term in Eq. \eqref{dispersion relation} will be dominant. For $k<k_c$ where the first term dominates, $l_{eff}=c_s/aH$, while for $k>k_c$ where the second term dominates, $l_{eff}=\sqrt[4]{\gamma}/aH$, so for the modes whose wavelength crosses the critical wavelength during inflation, the effective horizon will be discontinuous at the crossing point (denoted as $\tau_\ast$). This is different in normal inflation models without the $k^4$ correction term in the dispersion relation. However in this case, although the modified dispersion relation can affect the initial condition of the fluctuations, at late times (especially the superhorizon region) it can hardly make any effect on the perturbations so as to deviate from the standard slow-roll inflation. Therefore, the PBHs are not easy to be generated, either. 

If we break the “slow-varying” approximations of some of the variables, however, things will become different. Note that the critical scale $k_c$ is related to the sound speed $c_s$, and if we make $c_s$ steplike from a large value to a small value, $k_c$ will be large at the beginning, then become small later. For certain fluctuation modes with wave number $k$, one can have $k<k_c$ at the beginning, then $k>k_c$ later. We draw the same plot for this case in the right panel of Fig.~\ref{shiyitu}. In this case, the dispersion relation will be dominated first by the $k^2$ term, then the $k^4$ term. Moreover, the effective horizon will also be made steplike, namely $c_s/aH$ followed by $\sqrt[4]{\gamma}/aH$. Note that although for large-scale modes which exit the horizon before the transition time $\tau_\ast$, the $k^4$ term actually does not affect the solution (since $z^{\prime\prime}/z>\omega^2$), for small-scale modes which exit the horizon after the transition time, the $k^4$ term does affect the solution before the horizon crossing ($z^{\prime\prime}/z<\omega^2$). Therefore, PBHs can be formed in such a case, as has been shown in Ref. \cite{Ballesteros:2021fsp,Gorji:2021isn,Zhai:2022mpi} as well. We will analyze such a case in a bit more detail in the next section.

\section{our model with varying sound of speed}\label{model building} 

First of all, we assume that the inflation field $\phi$ still obeys the “slow-roll” approximation, namely

\begin{align}
	\left|\ddot{\phi} \right|\ll \left|3H\dot{\phi}\right|
	\,,\quad
	\frac{\left|2\beta \dot{H}\right|}{\left|\alpha M^{2}+3\beta H^{2}\right|} \ll 1
	\, ,\quad x_\beta\ll 1\, ,\quad y\ll 1,\quad {\cal D}\simeq 1\, ,
\end{align}
under which the equation of motion~\eqref{eom} and Friedmann equation are reduced to
\begin{align}
\label{friedmann}
	3\left(\alpha +\frac{3\beta H^{2}}{M^{2}}\right)H\dot{\phi}+V_{\phi} \simeq 0
	\,,\quad
	\frac{3 H^{2}}{\kappa^{2}} \simeq V(\phi)+\left(\alpha +\frac{6\beta H^{2}}{M^{2}}\right)\dot{\phi}^{2}
	.
\end{align}
In this case, the model is approaching a potential-driven inflation model. Thus the slow-roll parameter $\epsilon$ can be expressed in terms of the potential, namely $\epsilon=V_{\phi}^{2}/\left(2\kappa^{2}V(\phi)^{2}\right)$, while from the previous section, the sound speed squared $c_s^2$ can be expressed as
\begin{align}
\label{cs2}
c_s^2=\frac{\epsilon{\cal D}}{3x_\beta}\simeq \frac{1}{6\kappa^2x_\beta}\left(\frac{V_{\phi}}{V(\phi)}\right)^{2},
\end{align}
Therefore both $\epsilon$ and $c_s^2$ are closely related to the form of the potential $V(\phi)$.

According to the analysis above, we now consider a steplike sound speed form. One such parametrization is as the following function:
\begin{align}
    c_{s}^{2} = \frac{1}{A+ \exp[B\left(t-t_{c}\right)/\kappa]},
    \label{cs2-2}
\end{align}
where $A$, $B$ are parameters, and $t_{c}$ denotes the transition time. Therefore, when $t\ll t_c$, the second term in the dominator of Eq.\eqref{cs2-2} is suppressed exponentially, therefore we have $c_s^2\simeq 1/A$, which can be viewed as the initial value of $c_s^2$. On the other hand, when $t\gg t_c$, Eq.\eqref{cs2-2} as a whole is suppressed by the exponential term, giving rise to $c_s^2\rightarrow 0$. 
 
Making use of Eqs. \eqref{xandy}, \eqref{friedmann}, \eqref{cs2} and \eqref{cs2-2}, one can get the forms of potential $V(\phi)$ and function $f(\phi)$ as
\begin{eqnarray}
    V(\phi) &=& V_{0} \exp \left[-\frac{\sqrt{3} M\kappa}{\sqrt{A}B C} \tanh ^{-1}\sqrt{1+A^{-1}e^{B \left(\sqrt{2 C} \phi/M-t_{c}/\kappa\right)}}\right],
    \label{potential}\\
    f(\phi) &=& 72 C^{3} \gamma\left(\frac{\kappa}{V (\phi) M}\right)^{2/3} \left[A+e^{B \left(\sqrt{2 C} \phi/M-t_{c}/\kappa\right)}\right]^{1/3},
   \label{fphi}
\end{eqnarray}

and an analytical solution:
\begin{align}
	\phi(t) =\frac{1}{\sqrt{2 C}}\frac{M}{\kappa} t.
\end{align}
Here we also assume $\gamma$ is a constant.

\begin{figure*}[htbp]
	\centering
	\subfigure{
		\includegraphics[height=4.0cm,width=6.5cm]{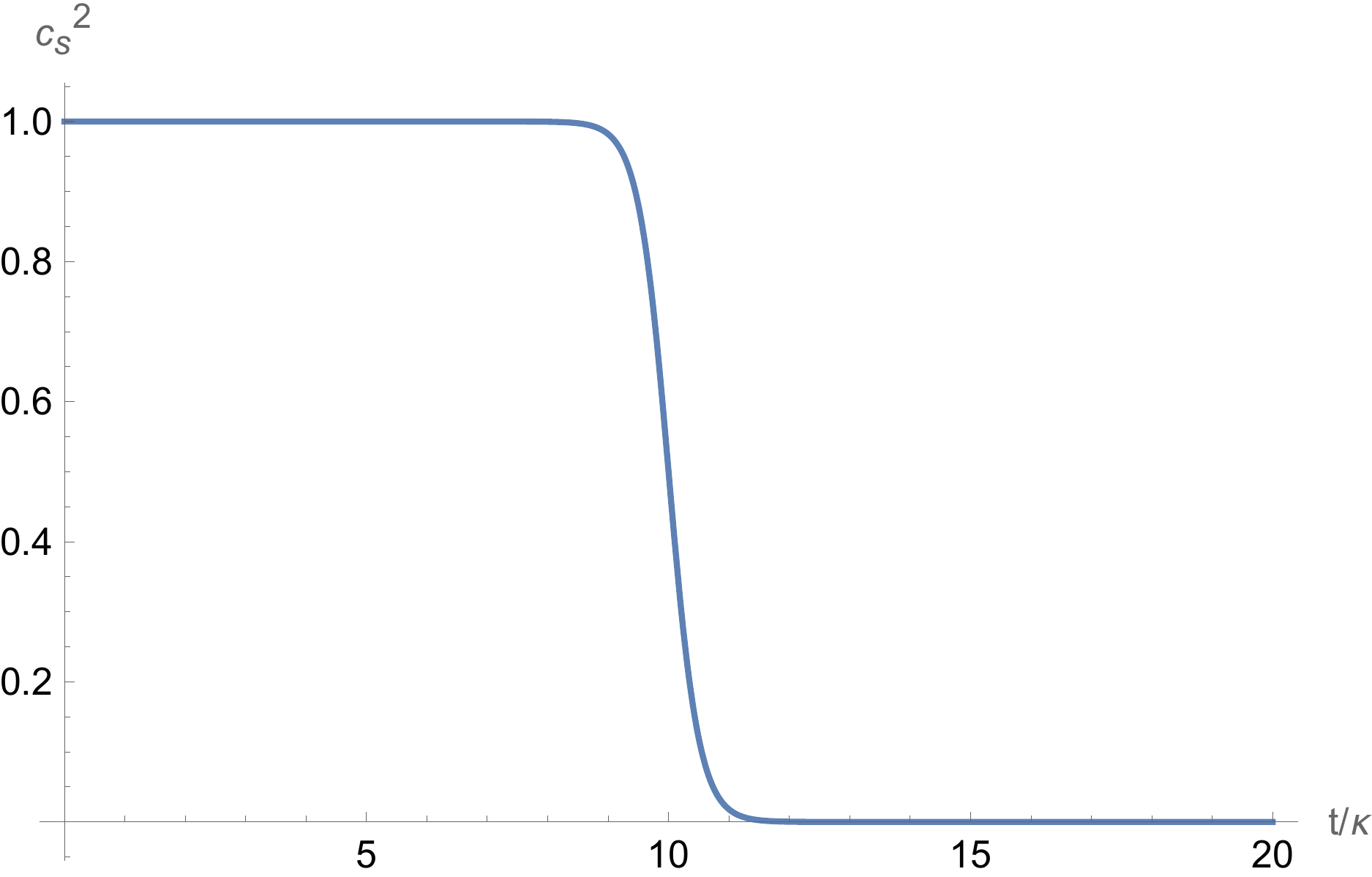}
	}
	\subfigure{
		\includegraphics[height=4.0cm,width=6.5cm]{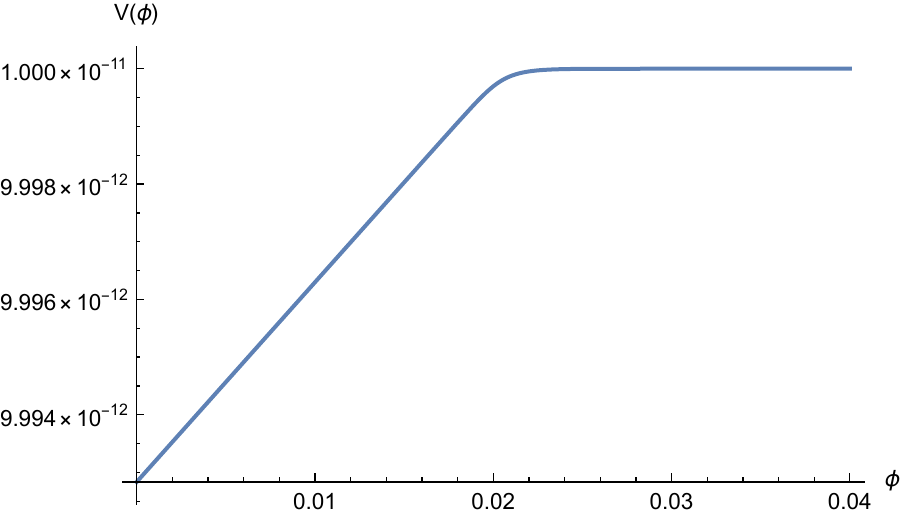}
	}
	\subfigure{
		\includegraphics[height=4.0cm,width=6.5cm]{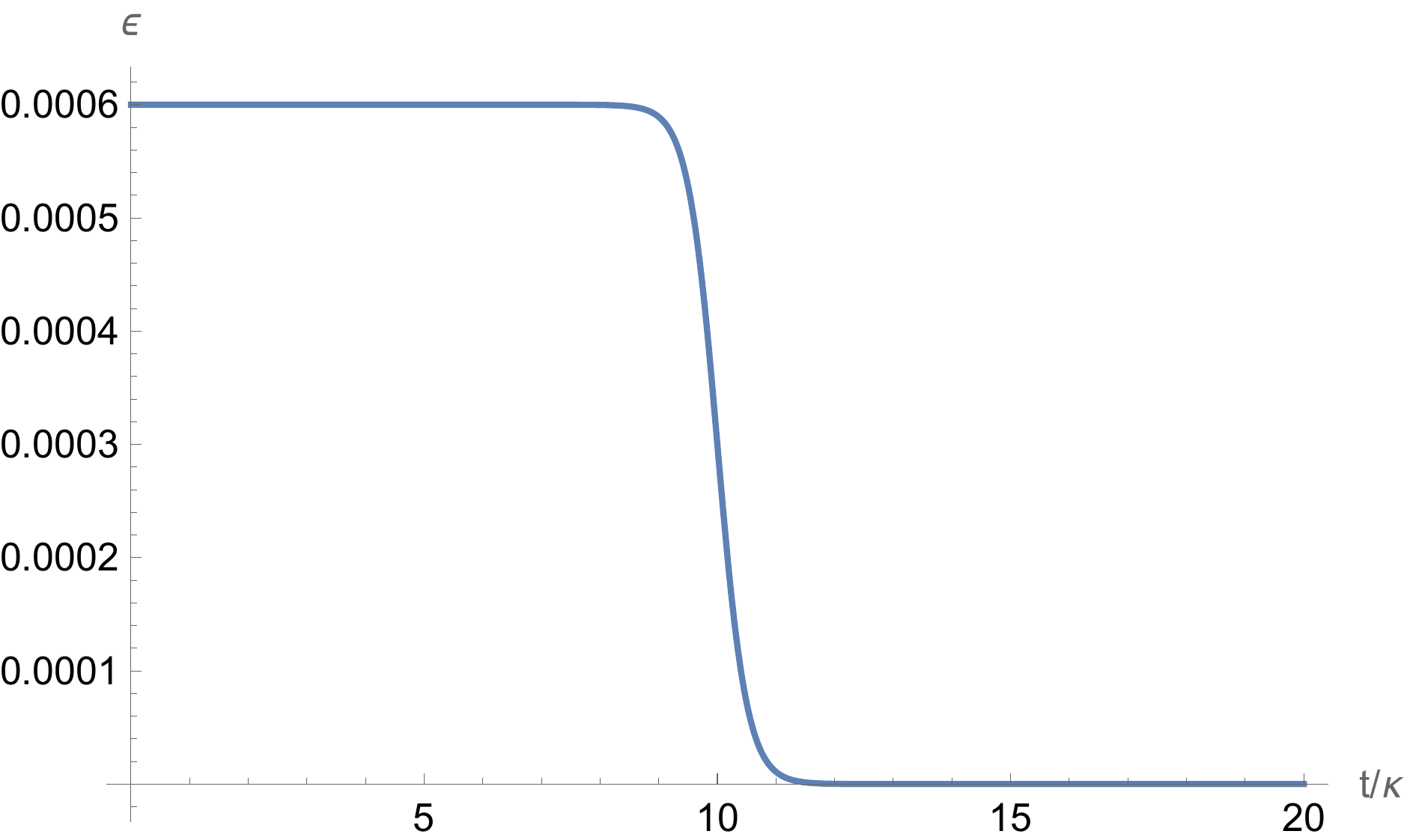}
	}
	
\caption{The evolutions of sound speed $c_s^2$, potential $V(\phi)$ and slow-roll parameter $\epsilon$ during inflation, whose analytical expressions are given by Eqs. \eqref{cs2-2}, \eqref{potential} and \eqref{fphi}, respectively. The related parameters are chosen as $A=1$, $B= 4$, $C=1250$, ${t}_{c}=10 M_{pl}^{-1}$, $M = 10^{-1} M_{pl}$, $V_{0} = 1\times 10^{-11} M_{pl}^{4}$.}
	\label{FIG3}
\end{figure*}

In Fig. \ref{FIG3}, we plot the evolution of $c_s^2$, $V(\phi)$ and $\epsilon$ in our model. We choose the parameters as 
$A=1$, $B= 4$, $C=1250$, ${t}_{c}=10 M_{pl}^{-1}$, $M = 10^{-1} M_{pl}$, $V_{0} = 1\times 10^{-11} M_{pl}^{4}$. The plot shows a sudden decrease of $c_s$ and $\epsilon$ at the middle stage of inflation, and such decrease is simultaneous, as can be seen  from Eq. \eqref{cs2}. Therefore in our model actually both $c_s$ and $\epsilon$ will contribute to the increase in the power spectrum. This is different from the discussions in \cite{Ballesteros:2021fsp,Gorji:2021isn,Zhai:2022mpi}.

As a side remark, let us mention that it is hard to get the simple mathematical forms of potential and functions simultaneously due to the nonlinear term in the action. However, the most important is the above calculation results show a steplike sound speed, which can generate PBHs as we want. Meanwhile, although the concise form of sound speed leads to a complex mathematical form of the potential, it can be seen from Fig.~\ref{FIG3} that the potential is a flat potential; in this sense, the potential form is simple and natural. It makes sense to optimize the model by finding a more concise mathematical expression for each function in the model, and this will be a step for future research.

\section{The evolution of the perturbations}\label{analytical solution}
In this section, we will calculate the analytical solution of Eq.~\eqref{perturbeomscalar2} based on the relationship between $k$ and $k_{c}$. In order to solve Eq.~\eqref{perturbeomscalar2}, we assume the $c_{s}$, $\epsilon$, $x_{\beta}$ and $y$ do not change significantly in both the $k<k_c$ and the $k>k_c$ region, and we find analytical solutions for each region. Furthermore, we match the solutions in the two regions by making use of an appropriate matching condition.
	
Let us first consider the $k<k_c$ region, where the $k^{2}$ term dominates over the $k^4$ term in the dispersion relation \eqref{dispersion relation}. Then Eq.~\eqref{perturbeomscalar2} reduces to
\begin{align}\label{k^2}
   u^{(-)\prime\prime}_{k}+\left(c_{s}^{2}k^{2}-\frac{2}{\tau^{2}} \right)u^{(-)}_{k}=0~. 
\end{align}
where $(-)$ denotes the variables in the $k<k_c$ region. Imposing the Bunch-Davies initial condition, we find the positive frequency solution for Eq.~\eqref{k^2} as
	\begin{align}\label{k^21}
		u^{(-)}_{k}=e^{i\pi}\frac{\sqrt{-\pi\tau}}{2}H_{3/2}^{(1)}(-c_{s}k\tau).
	\end{align}
We can obtain the power spectrum expression as follows
	\begin{align}
		P^{(-)}_{S}=\frac{k^{3}}{2\pi^{2}}\left|\zeta^{(-)} \right|^{2} =\frac{H^{2}\mathcal{D}}{24\pi^{2}c_{s}^{3}x_{\beta}}.
	\end{align}
Meanwhile, in the $k>k_c$ region where the $k^{4}$ term dominates over the $k^2$ term in \eqref{dispersion relation}, Eq.~\eqref{perturbeomscalar2} reduces to
	\begin{align}\label{Eq.k^4}
		u^{(+)\prime\prime}_{k}+\left(\frac{8}{3}x_{\beta}^{3}y k^{4}\tau^{2}-\frac{2}{\tau^{2}} \right)u^{(+)}_{k}=0,
	\end{align}
where $(+)$ denotes the variables in the $k>k_c$ region. The most general solution of Eq.~\eqref{Eq.k^4} can be written as
	\begin{align}\label{k^4}
		u^{(+)}_{k}=\frac{\sqrt{-\pi\tau}}{2}\left[C_1H_{3/4}^{(1)}(\frac{1}{2}\sqrt{\gamma}k^{2}\tau^{2})+C_2H_{3/4}^{(2)}(\frac{1}{2}\sqrt{\gamma}k^{2}\tau^{2}) \right],
	\end{align}
where $C_1$ and $C_2$ are constants.

Matching the two solutions \eqref{k^21} and \eqref{k^4} at the conformal time $\tau_{\ast}$ so as to make the solutions in $k>k_c$ and $k<k_c$ regions continuous at the time point $\tau_\ast$ \cite{Gorji:2021isn}, we can get
\begin{align}
    C_1 &= \frac{\pi}{4}\left[2y_{c}H_{3/2}^{(1)}(x_{c}) H_{-1/4}^{(2)}(y_{c})- x_{c}H_{1/2}^{(1)}(x_{c}) H_{3/4}^{(2)}(y_{c})  \right],
    \\
    C_2 &= - \frac{\pi}{4}\left[2y_{c}H_{3/2}^{(1)}(x_{c}) H_{-1/4}^{(1)}(y_{c})- x_{c}H_{1/2}^{(1)}(x_{c}) H_{3/4}^{(1)}(y_{c})  \right],
\end{align}
therefore
	\begin{align}\label{C1}
		\left|C_1-C_2 \right|^{2} &= 
		-\frac{\pi }{\sqrt{2}\left(-x_{c} \right)^{3}y_{c}^{3/2}}
		\left \{4HF[-\frac{3}{4},-\frac{1}{4}y_{c}^{2}]\left(x_{c}\cos(-x_{c})+\sin(-x_{c})\right)
		\notag \right. \\ 
		&\phantom{=\;\;}\left.
		+HF[\frac{1}{4},-\frac{1}{4}y_{c}^{2}]
		\left[3\left(x_{c}\cos(-x_{c})+\sin(-x_{c})\right)-x_{c}^{2}\sin(-x_{c}) \right] \right\}^{2}~,
	\end{align}
where $HF$ is the hypergeometric function, and we defined $x_{c}=-c_{s}k\tau$, $y_{c}=\frac{1}{2}\sqrt{\gamma}k^{2}\tau^{2}$.
	
For the fluctuation modes which exit the horizon at this stage, the power spectrum is as follows:
	\begin{align}\label{k^{4}}
		P^{(+)}_{S}=\frac{H^{2}}{4\pi}\left(\frac{16}{\gamma} \right)^{3/4}\frac{\mathcal{D}}{3x_{\beta}} \left(\Gamma(3/4) \right)^{2}\left|C_1-C_2 \right|^{2}.
	\end{align}
Since $x_{c} \gg 1$ in the $k^{4}$ phase, and we consider that $HF[-\frac{3}{4},-\frac{1}{4}y_{c}^{2}]\simeq \Gamma(1/4)^{-1}$, then Eq.~\eqref{C1} can be reduced to
	\begin{align}
		\left|C_1-C_2 \right|^{2} &\simeq \frac{\pi}{\sqrt{2}\left(x_{c} \right)^{3}y_{c}^{3/2}}\frac{x_{c}^{4}}{\Gamma(1/4)^{2}},
	\end{align}
at the transition point we can find
	\begin{align}\label{yc}
		y_{c} = \frac{x_{c}^{2}}{c_{s,t}^{2}}\left(\frac{2}{3}x_{\beta}^{3}y\right)^{1/2},
	\end{align}
at the same time,we consider $y_{c}\simeq 1$. Then substituting this value into Eq.~\eqref{k^{4}}, we find
	\begin{align}
 \label{spectrum}
		P^{(+)}_{S} = \frac{16\pi^{2}c_{s}^{3}c_{s,t}}{\gamma}\left(\frac{\Gamma(\frac{3}{4})}{\Gamma(\frac{1}{4})}\right)^{2} P^{(-)}_{S}\left(k \to 0\right),
	\end{align}
where $P^{(-)}_{S}\left(k \to 0\right) \simeq 2\times 10^{-9}$ is the power spectrum of fluctuation which exits the horizon before transition time $\tau_\ast$, thus constrained by the CMB measurements \cite{Planck:2018jri}. $c_{s}$ is the sound speed in the $k^{2}$ phase before the sound speed starts changing in time, $c_{s,t}$ is the sound speed at the transition point. For enough formation of PBHs, one needs the power spectrum up to $10^{-2}$~\cite{Motohashi:2017kbs}, therefore $\gamma \simeq 10^{-6}$ is required. A more precise numerical calculation shows that we have to assume $\gamma = 10^{-7}$. The above result is the same as Refs.~\cite{Ballesteros:2021fsp,Gorji:2021isn}.
	
\begin{figure*}
		\centering
		\includegraphics[height=5cm,width=8cm]{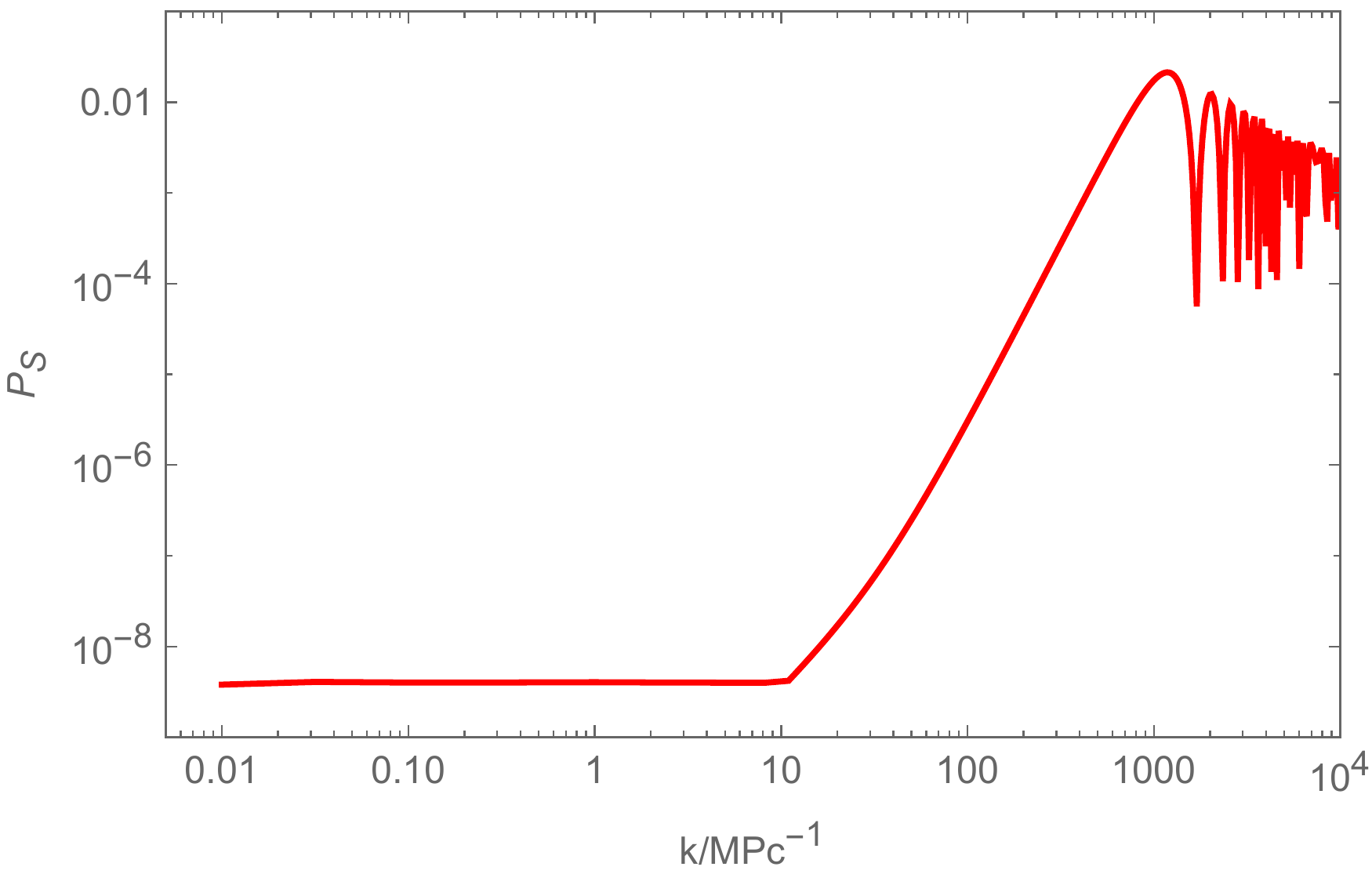}
		\caption{The plot of power spectrum in our model in terms of $k$. For $k<10 \text{Mpc}^{-1}$, the spectrum is flat and the amplitude fits with the CMB observational data. For $10 \text{Mpc}^{-1}<k<10^3 \text{Mpc}^{-1}$, the power spectrum increases with $k$ and reaches a peak at around $k\sim 10^3 \text{Mpc}^{-1}$. For $k>10^3 \text{Mpc}^{-1}$ where the fluctuation modes go deep into the subhorizon, the spectrum exhibits oscillating behavior. } 
		\label{ps1}
\end{figure*}

In the Fig.~\ref{ps1}, we show the relationship between power spectrum $P_S$ and the wavenumber $k$. From the plot we can see that, for the small $k$ region where $k<10\text{Mpc}^{-1}$, the fluctuations exit the horizon before the transition time, therefore the power spectrum is not affected by the decrease of the sound speed. Therefore, the amplitude of the power spectrum remains $\sim 10^{-9}$, consistent with the CMB observational data \cite{Planck:2018jri}. For the large $k$ region, the fluctuations exit the horizon after the decrease of sound speed, thus the amplitude of the power spectrum gets enhanced. The peak of the spectrum at the point $k\simeq 10^3 
 \text{Mpc}^{-1}$, corresponding to the scale where PBHs are generated. For the very large $k$ region where $k>10^3 \text{Mpc}^{-1}$, since the fluctuation modes are mainly in the subhorizon region, the oscillation behavior is robust.

\section{The abundance of Primordial Black Holes} \label{PBHs}

In this section, we consider the formation of the primordial black holes (PBHs) and their abundance in our model. Generally, after the end of inflation, our Universe will enter a radiation-dominated era, and the perturbations will reenter the horizon. If the perturbations are still very large, it will generate the primordial black hole by means of the gravitational collapse of the local inhomogeneities. From \cite{Green:1997sz,Inomata:2017okj,Sasaki:2018dmp}, the mass of PBHs formed during the radiation-dominated period can be described by the following equation:
\begin{align}
		M_{PBH}=\Upsilon M_{H}&=\Upsilon \frac{4\pi}{3}\rho_{form} H_{form}^{-3}
		\nonumber\\	
		&=10^{15}\times \left(\frac{\Upsilon}{0.2}\right)\left(\frac{g_{\star}}{106.75}\right)^{-1/6}\left(\frac{k_{form}}{0.07 Mpc^{-1}}\right)^{-2} M_{\odot},       
\end{align}
where $g_{\star}$ is the total effective degree of freedom of the Universe, and $\Upsilon \simeq 0.2$ is the efficiency factor; both are evaluated in the radiation dominated era~\cite{Sasaki:2018dmp}, and $M_{\odot}$ is the solar mass. $k_{form}$ is the comoving number of the fluctuations which formed PBHs, whose inverse denotes the scale of PBH formation. From the above expression we can see that the mass of the primordial black hole is determined by $k_{form}$. 

From the analysis in the above section, $k_{form}\simeq 1.174\times 10^3 \text{Mpc}^{-1}$. Compared to the wave number corresponding to solar-mass PBHs which is $\sim 10^5 \text{Mpc}^{-1}$, in our model the fluctuations exit the horizon not far from the window opened for CMB observations, and will reenter the horizon later than those of solar-mass. In this case, the PBHs have the opportunity to accumulate utill they became very massive. From the above equation we can get that, for our case with $k_{form}\sim {\cal O}(10^3) \text{Mpc}^{-1}$, the mass of PBH is around $M_{PBH}\sim 10^6-10^7 M_{\odot}$. Such a massive PBH can also help explain the generation of supermassive black holes, such as those with a mass of $10^{10}M_{\odot}$ found by a redshift of $z\simeq 7$ \cite{Wu_2015,Yang_2020,Carr:2020erq,Yang:2021imt}.

In order to evaluate how much primordial black holes can be generated and how it can act as the dark matter, we usually define the fraction of primordial black holes in dark matter namely, 
\begin{align}
\label{fpbh}
		f_{PBH}&=\left. \beta(M_{PBH})\left(\frac{\rho_{tot}}{\rho _{DM}}\right)\right|_{form}
		\nonumber\\
		&=1.68\times 10^{8}(\frac{\Upsilon}{0.2})^{1/2} (\frac{g_{\star}}{106.75})^{-1/4}(\frac{M_{PBH}}{M_{\odot}})^{-1/2} \beta(M_{PBH}).	
\end{align}
where 
\begin{align}
    \beta(M_{PBH})=\left. \frac{\rho_{PBH}}{\rho_{tot}}\right|_{form} 
\end{align}
is the fraction of the PBHs in the entire Universe. The last step of \eqref{fpbh} comes from some tedious but straightforward calculation \cite{Sasaki:2018dmp,Zheng:2021vda}. On the other hand, according to the Press-Schechter formalism \cite{Press:1973iz,Green:2004wb}, $\beta$ is given by the probability that  the fractional overdensity $\delta\equiv\delta\rho/\rho$ is above a certain threshold $\delta_c$ for PBH formation \cite{Mahbub:2020row}. Therefore for Gaussian primordial fluctuations, $\beta$ is given by
\begin{align}
		\beta(M_{PBH})&=2 \int_{\delta_{c}}^{\infty} \exp\left(-\frac{\delta^{2}}{\sigma^{2}(M(k))}\right) \frac{d\delta}{\sqrt{2\pi}\sigma(M(k))}
		\nonumber\\
		&=\sqrt{\frac{2}{\pi}}\frac{\sigma(M(k))}{\delta_{c}}\exp\left(-\frac{\delta_{c}^{2}}{\sigma^{2}(M(k))}\right),
\end{align}
where $\delta_{c}$ is the threshold density. Here $\sigma^{2}(M(k))$  represents the standard deviation of the coarse-grained density contrast for the PBHs mass of $M$~\cite{Young:2014ana}:
\begin{align}
		\sigma^2\left( M\left( k \right) \right) =\frac{16}{81}\int_0^{\infty}{d\ln q}\left( \frac{q}{k} \right) ^4W\left( \frac{q}{k} \right) ^2P _{S}\left( q \right).
\end{align}
Therefore it can be related to the primordial power spectrum at the horizon reentering. In this work, we adopt the Gaussian window $W(x)=\exp \left( -x^2/2 \right)$. The result of the power spectrum is given by Eq. \eqref{spectrum} and, substituting it into the above formulas, one can get the fraction of PBHs generated in our model.

\begin{figure*}[htbp]
		\centering
             \includegraphics[height=5cm,width=8cm]{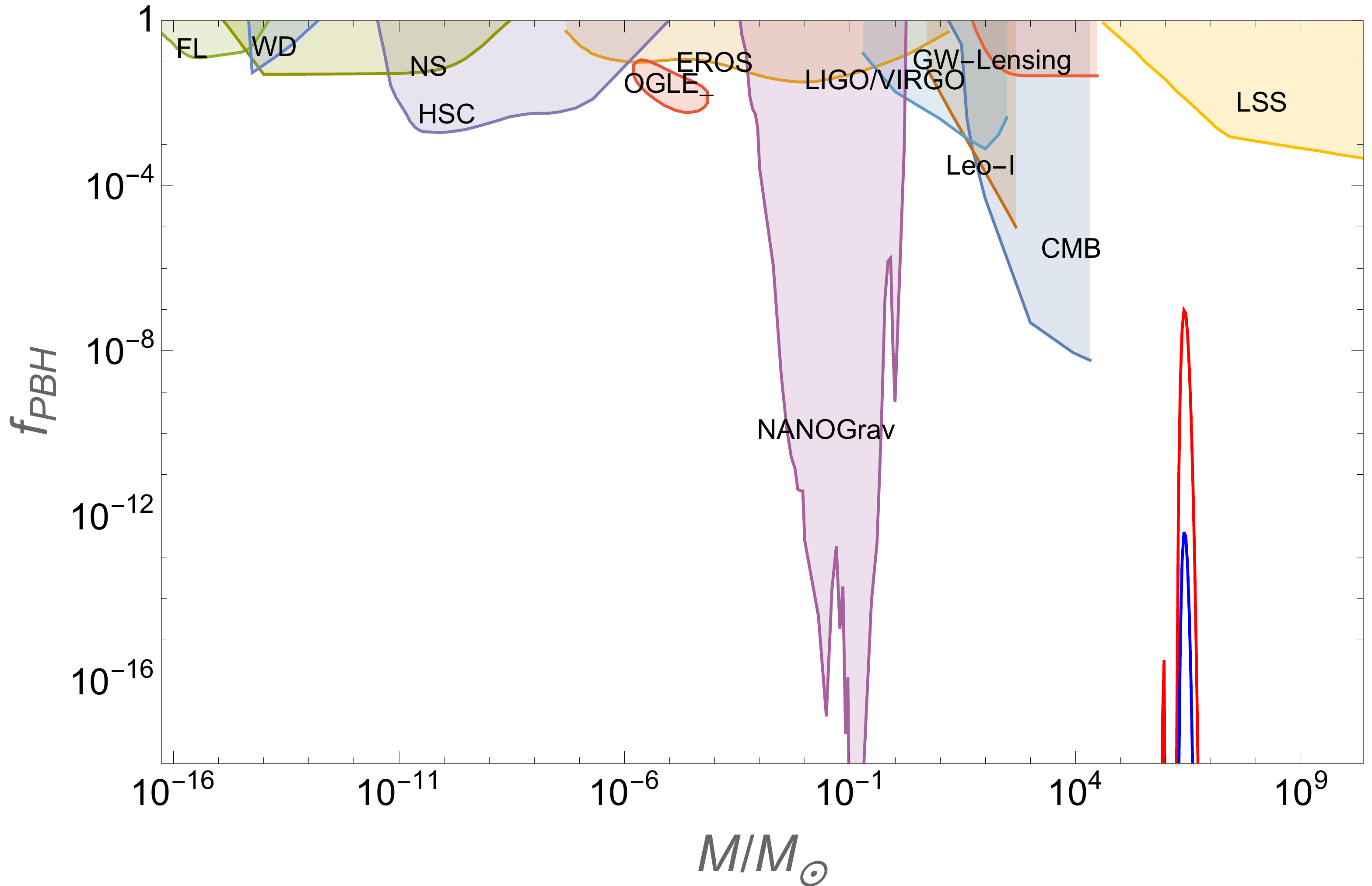}
		\caption{Colored peaks: the fraction of PBHs generated by our model against the total amount of dark matter. The parameters are chosen to be the same as before. The red peaks correspond to the threshold density $\delta_{c}=0.33$ which was initially obtained by Carr {$\it et al$}. using the relation $\delta_c\simeq w$ in the radiation-dominated era \cite{Carr:1975qj}, while the blue peaks correspond to $\delta_{c}=0.40$, which was obtained after a refined analytical and numerical calculation recently \cite{Harada:2013epa, Musco:2018rwt, Escriva:2019phb, Escriva:2020tak, Musco:2020jjb}. The highest peaks locates around $M/M_{\odot}\sim 10^7$, and reaches the value of $f_{PBH}\simeq 10^{-7}$. Colored regions: the excluded regions by various observations, which are obtained from the publicly available {\bf Python} code \href{https://github.com/bradkav/PBHbounds}{$\bf PBH_{BOINDS}$} \cite{bradley_j_kavanagh_2019_3538999}.}
		\label{Fig4}
\end{figure*}

In Fig.~\ref{Fig4}, we plot $f_{PBH}$ against the mass of PBHs, $M_{PBH}$, and confront various constraints that are obtained from the publicly available {\bf Python} code \href{https://github.com/bradkav/PBHbounds}{$\bf PBH_{BOUNDS}$} \cite{bradley_j_kavanagh_2019_3538999}. The constraints contain Experience de Recherche d’Objets Sombres (EROS)~\cite{EROS-2:2006ryy}, Subaru Hyper Suprime-Cam (Subaru-HSC)~\cite{Niikura:2017zjd}, Gravitational-Wave Lensing~\cite{Jung:2017flg}, Optical Gravitational Lensing Experiment (OGLE)~\cite{Niikura:2019kqi}, cosmic microwave background (CMB)~\cite{Serpico:2020ehh}, femtolensing of Gamma-ray bursts (FL)~\cite{Barnacka:2012bm}, white dwarf explosions (WD)~\cite{Graham:2015apa}, neutron stars (NSs)~\cite{Capela:2013yf} (note that it has later been shown that the survival of stars actually cannot constrain the PBHs, thus making the constraints even looser, see~\cite{Katz:2018zrn, Montero-Camacho:2019jte}), Leo-I dwarf galaxy~\cite{Lu:2019ktw}, NANOGrav~\cite{Chen:2019xse, Wong:2020yig}, LIGO/VIRGO~\cite{Kavanagh:2018ggo}, various cosmic large-scale structures $LSS_{s}$~\cite{Carr:2018rid} and so on. As demonstrated before, the mass range of the PBHs formed in our model is around $10^{6}-10^{7} M_{\odot}$. Moreover, we find that as the threshold density $\delta_{c}$ increases, the corresponding PBH abundance will decrease. This is easy to understand: the higher the threshold energy, the more difficult it is to form a black hole. The oscillating behavior of the power spectrum can also lead to the generation of multimass PBHs.
	
\section{Scalar Induced Gravitational Waves} \label{SIGWs}
The large amount of the primordial scalar perturbations cannot only generate primordial black holes via overdensity collapse, but also induce gravitational waves in the radiation-dominated era ~\cite{Ananda:2006af,Baumann:2007zm}. Such kinds of gravitational waves are effects of second order, thus are usually neglected in large scales such as CMB scale, where the scalar perturbation source are already constrained to be small. However, in small scales, the scalar induced gravitational waves (SIGWs) may become large to be detected, and thus become a probe to small scale physics as well.  	
The energy densities of SIGWs at present are related to their values after the horizon reentry in the RD era as~\cite{Pi:2020otn,Chen:2021nio}
\begin{align}\label{eq:GWES}
		\Omega_{GW,0}(k) h^2 =  0.83\,\Omega_{RD, 0}h^2 \left( \frac{g_{c}}{10.75}\right)^{-\frac13}\Omega_{GW}(k, \eta_c)~,
\end{align}	
where $\Omega_{RD,0} h^2\simeq 4.2\times 10^{-5}$ 
is the current radiation density parameter and the $g_{c} \simeq 106.75$ is the effective degrees of freedom in the energy density at $\eta_{c}$, at which $\Omega_{GW}$ stops growing. The energy density of the GWs in the radiation-dominated era is~\cite{Kohri:2018awv,Cai:2018dig,Fu:2019vqc,Lu:2019sti}
\begin{align}
		\Omega_{GW}(k,\eta_{c}) = \frac{1}{6}  \int^\infty_0 dv \int^{1+v}_{|1-v|} du \left[ \frac{4v^2-(1-u^2+v^2)}{4uv}\right]^2 \overline{I_{RD}^2(u,v)} \mathcal{P}_S(kv)\mathcal{P}_S(ku),
\end{align}
where the variables $u$ and $v$ are defined as $u\equiv |\textbf{k}-\tilde{\textbf{k}}|/k$, $v\equiv \tilde{k}/k$, and the full expression of $I_{RD}$ is given by
\begin{align}
		\overline{I_{RD}^2(u, v)} =& \frac{9}{u^2 v^2} \left(\frac{u^2+v^2-3}{2 u v}\right)^4\nonumber\\
		&\times \left[\left( \ln\left| \frac{3-(u+v)^2}{3-(u-v)^2}\right|-\frac{4 u v}{u^2+v^2-3}\right)^2+\pi ^2 \Theta \left(u+v-\sqrt{3}\right)\right],
\end{align}
$\Theta$ is the Heaviside theta function. 
Moreover, from the relationship of the frequency of gravitational waves $f$ and the wave number $k$,
\begin{align}
		f = 1.546\times 10^{-15}\left( \frac{k}{\text{Mpc}^{-1}}\right) 
 \text{Hz},
\end{align}
\begin{figure*}
		\centering
		\includegraphics[height=5cm,width=8cm]{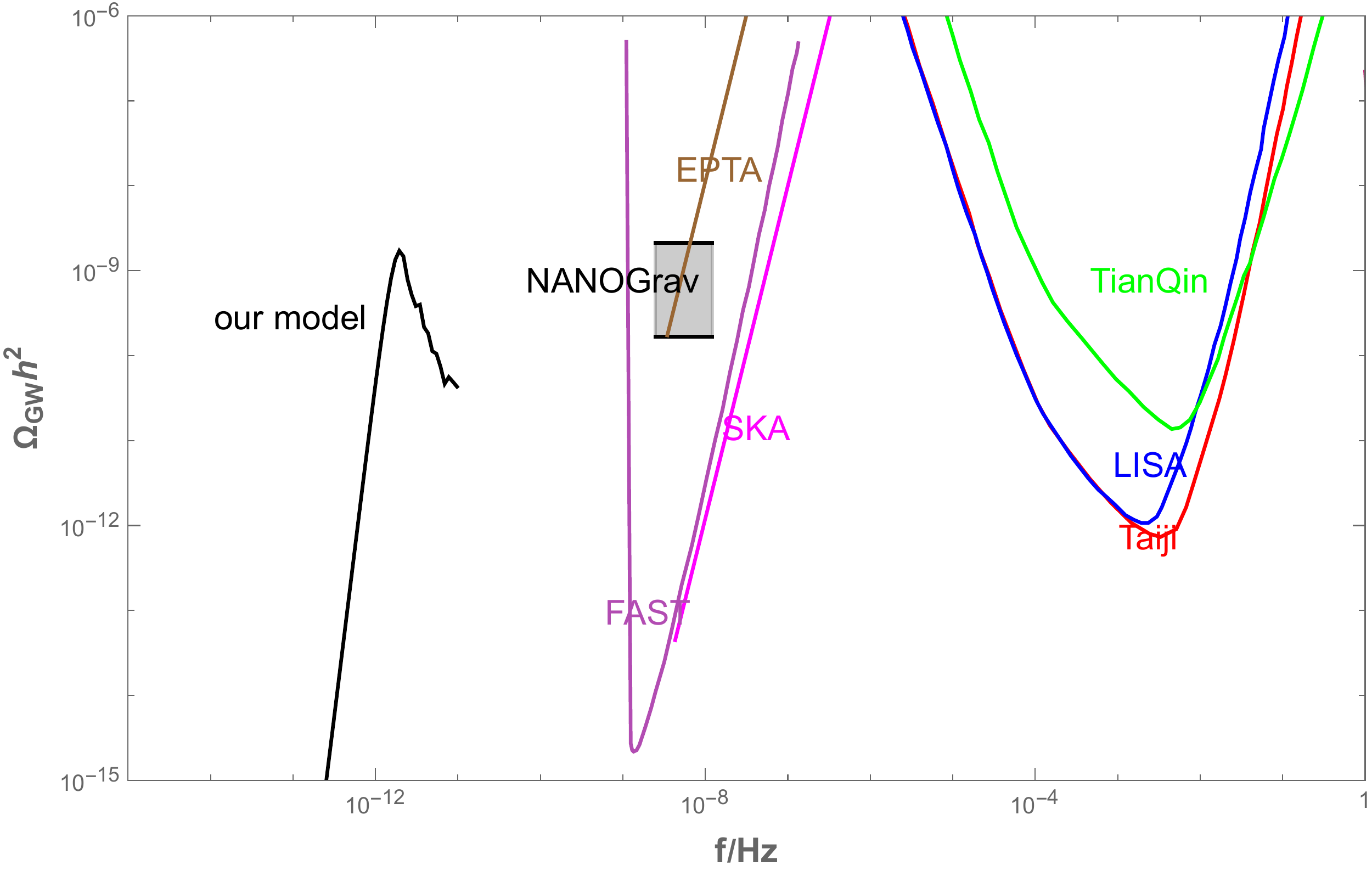}
		\caption{Black line: the energy spectra of SIGWs in our model. The parameters are chosen to be the same as before. Colored lines: the sensitivity curves of the current and future GW projects, including EPTA, NANOGrav (gray region), SKA, FAST, TianQin, LISA and Taiji.
		}
		\label{sigws}
\end{figure*}
we can easily have $f\simeq 10^{-12}$Hz for $k_{form}\simeq 10^{3} \text{Mpc}^{-1}$ in our model. In Fig.~\ref{sigws}, we plot the SIGWs generated in our model as well as the constraints from experiments such as EPTA~\cite{Lentati:2015qwp,Desvignes:2016yex}, NANOGrav~\cite{NANOGrav:2020bcs}, SKA~\cite{Janssen:2014dka}, FAST~\cite{Nan:2011um}, TianQin~\cite{TianQin:2015yph,TianQin:2020hid}, LISA~\cite{LISA:2017pwj}, and
Taiji~\cite{Hu:2017mde}. We can see that the current observations cannot give constraints to such a low frequency, however, since it is close to the primordial gravitational waves generated by quantum fluctuations in CMB scale, it is possible to have it detected in the upcoming CMB telescopes, such as AliCPT \cite{Li:2017drr}, or CMB-S4 \cite{CMB-S4:2016ple} collaborations. 
	
\section{Conclusions and Discussion} \label{Conclusion}
The generation of PBHs has attracted attentions from many people both in theoretical physics and astronomy, and has been widely investigated in recent years. To generate PBHs in the inflation era, one usually requires the primordial scalar perturbations increase and form a peak in the small scales, and this can be realized by the suppression of either the slow-roll parameter $\epsilon$, or the sound speed $c_s$, or maybe both. However, in order to decrease the sound speed, a modified dispersion relation of perturbations with higher order terms is needed in order not to break the effective field theory description of our Universe. A typical example of such a modification is to add a quadratic correction term, namely $\omega^2=c_s^2k^2+\alpha k^4$. 
	
In this work, we consider a specific inflation model where a nonminimal kinetic coupling term resides in the DBI-like square root in the action, which makes the action nonlinear, giving rise to a $k^{4}$ term in the equation as well as the dispersion relation of the scalar perturbation. Based on this model, we consider the sound speed that can vary from the early to late time during the inflation. While at the very beginning when $c_s\simeq 1$, the $k^2$ term dominates the dispersion relation, at the late time $c_s$ drops to a vanishing value, $k^2$ becomes negligible while the $k^4$ term becomes dominant. We constructed the potential for this model according to the sound speed. We calculated the evolution of perturbations during the whole region, and obtained the final power spectrum in both CMB scale and PBH formation scale. While in the CMB scale the spectrum is consistent with the observational constraint, in the PBH formation scale it has a peak of $10^{-2}$, which is sufficient to allow PBHs to form and act as the origin of dark matter in the Universe.

Making use of the Press-Schechter formalism, we calculate the mass of the generated PBHs $M_{PBH}$, as well as the fraction of PBHs against the total amount of dark matter, $f_{PBH}$. We found that the scale of the PBH formation is near to the CMB scale, which means that the PBHs in our model are formed later than the usual solar-mass ones, and more massive, falling into the category of supermassive black holes. The fraction of PBHs in our model is consistent with constraints by various observations, from the gravitational lensing to large scale structures. Moreover, we also investigated the scalar-induced gravitational waves, and find that the frequency of the gravitational waves is much lower than the current gravitational wave observations both using interferometers and pulsars, but close to that of the primordial gravitational waves generated from quantum fluctuations. 

Some final remarks are in order. First of all, from the theoretical point of view, we know that the PBH formation is a highly nonlinear process, which might cause large non-Gaussianities and backreactions such as loop corrections. It is still not clear whether these backreactions will affect (or ruin) the PBH formation process investigated here. Although the constraints from these effects are still not decisive yet (see recent discussions in \cite{Kristiano:2022maq}), we cannot say that it will not become a smoking gun in the future. Second, as mentioned before, the PBHs generated in our model are supermassive ones, different from the normal ones which have asteroid, lunar or solar masses. This may also be interesting to the studies in astrophysics, for it may be possible to explain the findings of supermassive quasars in $z\simeq 7$ as well as the galaxy formations. Moreover, the low frequency of scalar-induced gravitational waves may also attract the attention of the next generation of CMB and primordial gravitational wave detections. Making use of these detections, we may be able to test the model by verifying whether there is such a low frequency SIGW or not. We will extend these investigations in upcoming works.

\begin{acknowledgments}
We are grateful to Jiaming Shi for useful discussions. This work is supported by the National Key Research and Development Program of China under Grant NO. 2021YFC2203100, and the National Science Foundation of China under Grant No. 11875141. 
\end{acknowledgments}
	
\bibliographystyle{apsrev4-1}
\bibliography{References}	
\end{document}